\pdfoutput=1
\documentclass[11pt]{article}
\usepackage{jheppub_mod}
\usepackage{graphicx}
\usepackage{amsmath, amssymb}
\usepackage{bm}
\usepackage{mathrsfs}
\usepackage{empheq}
\usepackage{mathtools}
\usepackage{amsthm}
\usepackage{color}

\newtheorem{theorem}{Theorem}
\newtheorem{lemma}{Lemma}

\newtheorem{cor}{Corollary}
\theoremstyle{definition}
\newtheorem{definition}{Definition}
\theoremstyle{remark}
\newtheorem*{remark}{Remark}

\newcommand{\q}{{\mathsf q}}
\newcommand{\p}{{\mathsf p}}
\def\EXP{\textrm{{\large e}}}
\newcommand{\ii}{i}

\def\im{\mathop{\hbox{\rm Im}}\nolimits}

\newcommand{\veca}{\bm{a}}

\newcommand{\vect}{\bm{t}}
\newcommand{\vecu}{\bm{u}}
\newcommand{\vecb}{\bm{b}}

\newcommand{\vecz}{\bm{z}}
\newcommand{\vecy}{\bm{y}}
\newcommand{\cS}{\mathcal{S}}
\newcommand{\tN}{\widetilde{N}}

\title{Elliptic hypergeometric sum/integrals for orthogonal groups and Seiberg duality on the lens space}

\author[a]{Andrew P. Kels}
\author[b,c,d,e]{and Masahito Yamazaki}

\affiliation[a]{School of Mathematics and Statistics, The University of New South Wales,\\ Sydney, NSW 2052, Australia}
\affiliation[b]{Department of Physics, University of Tokyo, Tokyo 113-0033, Japan}
\affiliation[c]{Kavli Institute for Physics and Mathematics of the Universe (Kavli IPMU), UTIAS,\\ University of Tokyo, Chiba 277-8583, Japan}
\affiliation[d]{Trans-Scale Quantum Science Institute, University of Tokyo, Tokyo 113-0033, Japan}
\affiliation[e]{Center for Interdisciplinary Theoretical and Mathematical Sciences (iTHEMS), RIKEN,\\ Saitama 351-0198, Japan}

\preprint{RIKEN-iTHEMS-Report-26}

\abstract{We derive new transformation formulas for elliptic hypergeometric sum/integrals associated with the orthogonal groups $\mathrm{SO}(N)$. The sum/integrals are written in terms of the lens elliptic gamma function, which depends on an integer $r\geq 1$ in addition to the usual complex variables, and they reduce to the elliptic hypergeometric integrals of Spiridonov and Vartanov for $r=1$. The derivation uses the $BC_n\leftrightarrow BC_{\tilde n}$ transformation formula proven in our previous paper, together with a duplication formula for the lens elliptic gamma function proven here. The duplication formula takes a different form depending on whether $r$ is even or odd. Physically, the new transformations show that the supersymmetric lens indices on $S^1\times S^3/\mathbb{Z}_r$ of the 4d $\mathcal{N}=1$ $\mathrm{SO}(N)$ and $\mathrm{SO}(N_f-N+4)$ gauge theories related by Seiberg duality are equal for odd $r$. For even $r$, where the lens index depends on the global form of the gauge group, they show the equality of the contributions from flat connections whose holonomies lie in a common maximal torus. The $BC_n$ formula itself gives the corresponding statement for the symplectic groups.}

\begin{document}

\maketitle

\section{Introduction}

Supersymmetric gauge theories have been a rich source of identities for elliptic hypergeometric integrals. The superconformal index of a 4d $\mathcal{N}=1$ theory \cite{Romelsberger:2005eg,Kinney:2005ej} is given \cite{Romelsberger:2007ec,Dolan:2008qi} by an elliptic hypergeometric integral \cite{MR2479997}. Seiberg duality \cite{Seiberg:1994pq} predicts that the indices of the electric and magnetic theories are equal, and this equality is a transformation formula for the corresponding integrals. In this way the elliptic beta integral \cite{MR1846786} corresponds to the duality for $\mathrm{SU}(2)=\mathrm{Sp}(2)$ with six fundamental chiral multiplets, and Rains' transformation formulas for the $A_n$ and $BC_n$ root systems \cite{Rains:2005lkc} correspond to dualities with $\mathrm{SU}(N)$ and $\mathrm{Sp}(2N)$ gauge groups \cite{Dolan:2008qi,Spiridonov:2009za}. Spiridonov and Vartanov extended this dictionary to many other dualities \cite{Spiridonov:2009za,Spiridonov:2011hf}, including the Seiberg dualities for orthogonal gauge groups \cite{Intriligator:1995id}.

Another way to generalise these identities is to replace $S^1\times S^3$ by $S^1\times S^3/\mathbb{Z}_r$ \cite{Benini:2011nc,Razamat:2013opa}. The corresponding partition function, the lens index, contains a sum over holonomies around the non-contractible cycle of the lens space. Mathematically, the lens index is written in terms of the lens elliptic gamma function \cite{Benini:2011nc,Razamat:2013opa,Kels:2015bda,Gahramanov:2016ilb,Kels:2017toi}, and it takes the form of an elliptic hypergeometric \emph{sum/integral} (also called a rarefied elliptic hypergeometric integral \cite{Spiridonov:2016uae,Sarkissian:2018ppc}). The sum/integral analogue of the elliptic beta integral was proven in \cite{Kels:2015bda}, and $BC_n$ evaluation formulas were obtained in \cite{Spiridonov:2016uae}. In our previous paper \cite{Kels:2017toi} we proved sum/integral analogues of Rains' $A_n$ and $BC_n$ transformation formulas. These give the equality of lens indices for the $\mathrm{SU}(N)$ and $\mathrm{Sp}(2N)$ Seiberg dualities, and the $A_n$ formula also implies the star-star relation for the integrable lattice models of \cite{Yamazaki:2013nra}.

In this paper we complete this picture for the classical groups by treating the orthogonal groups $\mathrm{SO}(N)$, which were left open in \cite{Kels:2017toi}. Seiberg dualities for orthogonal gauge groups have been tested on the lens space by expanding the lens index for small $r$ \cite{Razamat:2013opa}. To our knowledge, the results of this paper give the first proof of the corresponding equalities of lens indices that is valid for all $r$; for even $r$ it concerns the sectors described below. Our main result, Theorem~\ref{thm:main}, is an equality between the $\mathrm{SO}(N)$ and $\mathrm{SO}(\tN)$ sum/integrals with $N+\tN=N_f+4$, valid for all $r$ and for every parity of $N$ and $\tN$. Following the strategy of \cite{Spiridonov:2011hf} for $r=1$, we obtain it from the $BC_n\leftrightarrow BC_{\tilde n}$ transformation of \cite{Kels:2017toi} by fixing six, seven or eight of the parameters to special values. The main new ingredient is a duplication formula for the lens elliptic gamma function (Lemma~\ref{lem:dup}). The special values in this formula differ for even and odd $r$. We choose them to be invariant under the duality map $t\mapsto (\sigma+\tau)/2-t$, $a\mapsto -a$. With this choice, all the extra factors produced along the way cancel, and the final identity has exactly the form physics predicts: the lens index of $\mathrm{SO}(N)$ SQCD equals that of the magnetic $\mathrm{SO}(N_f-N+4)$ theory with $N_f$ dual quarks, gauge singlet mesons in the symmetric tensor representation of the flavour group, and a cubic superpotential.

For odd $r$, Theorem~\ref{thm:main} is the equality of the full lens indices of the dual theories. For even $r$, the lens index depends on the global form of the gauge group and on discrete theta angles \cite{Razamat:2013opa,Aharony:2013hda,Razamat:2013jxa}, and it receives contributions from several topological sectors. Our sum/integrals capture the sectors in which the holonomies around both cycles of $S^1\times S^3/\mathbb{Z}_r$ lie in a common maximal torus, summed with equal weight. The remaining sectors, which distinguish $\mathrm{SO}(N)_+$, $\mathrm{SO}(N)_-$ and $\mathrm{Spin}(N)$, are not of this form, and we leave them for future work (see Section~\ref{sec:phys}).

The rest of the paper is organised as follows. Section~\ref{sec:defs} collects the definitions and properties of the lens elliptic gamma function. Section~\ref{sec:bcn} recalls the $BC_n\leftrightarrow BC_{\tilde n}$ transformation of \cite{Kels:2017toi} and its interpretation as $\mathrm{Sp}(2n)$ duality. Section~\ref{sec:dup} proves the duplication formula and related identities. Section~\ref{sec:so} defines the $\mathrm{SO}(N)$ sum/integrals and proves the main theorem. Section~\ref{sec:phys} discusses special cases and the physical interpretation. Section~\ref{sec:concl} contains concluding remarks.

\section{Lens elliptic gamma function}\label{sec:defs}

We use the conventions of \cite{Kels:2017toi}, which we briefly recall. Let $\sigma,\tau\in\mathbb{C}$ with
\begin{align}
\im(\sigma)\,,\; \im(\tau)>0\,,
\end{align}
and define the nomes
\begin{align}
\label{nomedef}
\p=\EXP^{2\pi\ii\sigma}\,,\qquad\q=\EXP^{2\pi\ii\tau}\,.
\end{align}
We also define
\begin{align}
\label{etadef}
\eta\coloneq \frac{\sigma+\tau}{2}\,.
\end{align}

Let $r$ be a positive integer.
The lens elliptic gamma function is \cite{Benini:2011nc,Razamat:2013opa}
\begin{align}
\label{legf2}
\Gamma(z,m;\sigma,\tau)\coloneq \EXP^{\phi_e(z, m)}\gamma(z,m)\,,\qquad z\in\mathbb{C}\,,\; m\in\mathbb{Z}\,,
\end{align}
where the normalisation is
\begin{align}
\label{ellnorm}
\begin{split}
\phi_e(z,m)\coloneq \frac{\pi\ii m(m-r)}{6r}\left(6(z-\eta)+(2m-r)(\sigma-\tau-1)\right)\,,
\end{split}
\end{align}
and
\begin{align}
\label{littlegamma}
\gamma(z,m)\coloneq \prod_{j,k=0}^\infty\frac{1-\EXP^{-2\pi\ii z}\p^{-m}(\p\q)^{j+1}\p^{r(k+1)}}{1-\EXP^{2\pi\ii z}\p^{m}(\p\q)^j\p^{rk}}\,\frac{1-\EXP^{-2\pi\ii z}\q^{-r+m}(\p\q)^{j+1}\q^{r(k+1)}}{1-\EXP^{2\pi\ii z}\q^{r-m}(\p\q)^j\q^{rk}}\,.
\end{align}
The normalisation \eqref{ellnorm} may also be expressed in terms of multiple Bernoulli polynomials \cite{Gahramanov:2016ilb,Kels:2017toi}.

For $r=1$, \eqref{legf2} is equivalent to the elliptic gamma function \cite{Ruijsenaars:1997aqs}, defined by
\begin{align}
\label{egf2}
\Gamma_{1}(z;\sigma,\tau)=\prod_{j,k=0}^\infty\frac{1-\EXP^{-2\pi\ii z}\p^{j+1}\q^{k+1}}{1-\EXP^{2\pi\ii z}\p^j \q^k}\,.
\end{align}
For general $r$, the infinite product $\gamma$ in \eqref{littlegamma} is a product of two elliptic gamma functions:
\begin{align}
\label{legfprod}
\gamma(z,m)=\Gamma_{1}(z+\sigma m;r\sigma,\sigma+\tau)\,\Gamma_{1}(z+\tau(r-m);r\tau,\sigma+\tau)\,.
\end{align}

Throughout the paper, unless we need to indicate the dependence on the nomes we will use the following shorthand notation for the lens elliptic gamma function
\begin{align}
    \Gamma(z,m)\coloneq \Gamma(z,m;\sigma,\tau)\,.
\end{align}
We will also use the following compact notations for products of the lens elliptic gamma function
\begin{align}
\label{compnot}
\begin{split}
\Gamma(t\pm z,a\pm y)&\coloneq \Gamma(t+z,a+y)\,\Gamma(t-z,a-y)\,, \\
\Gamma(\pm z,\pm y)&\coloneq \Gamma(z,y)\,\Gamma(-z,-y)\,, \\
\Gamma(\pm z_1\pm z_2,\pm y_1 \pm y_2)&\coloneq \Gamma(\pm(z_1+z_2),\pm(y_1+y_2))\,\Gamma(\pm(z_1-z_2),\pm(y_1-y_2))\,.
\end{split}
\end{align}
The lens elliptic gamma function satisfies the reflection formula
\begin{align}
\label{reflection}
\Gamma(2\eta-z,-m)\,\Gamma(z,m)=1\,.
\end{align}
In \cite{Kels:2017toi} the integer $m$ in \eqref{legf2} was restricted to $\{0,1,\ldots,r-1\}$, and $\Gamma(z,m)$ for other integers was defined through periodicity. This restriction is not needed, because the expression \eqref{legf2} is itself $r$-periodic in $m$. More precisely, the following periodicity and quasi-periodicity relations are satisfied
\begin{align}
\label{periodic}
\begin{split}
\Gamma(z,m+r)=&\,\Gamma(z,m)\,,\\ \Gamma(z+r,m)=&\,(-1)^{m(r+1)}\Gamma(z,m)\,,\\ \Gamma(z+1,m)=&\,\EXP^{\pi\ii\, m(m-r)/r}\,\Gamma(z,m)\,.
\end{split}
\end{align}
Indeed, for the first relation, we recall 
\begin{align}
\Gamma_1(X+\omega_1;\omega_1,\omega_2)=\theta(\EXP^{2\pi\ii X};\EXP^{2\pi\ii\omega_2})\,\Gamma_1(X;\omega_1,\omega_2)\,,
\end{align}
where 
\begin{align}
\theta(x;\q)=(x;\q)_\infty(\q/x;\q)_\infty\,, \qquad (x;\q)_\infty=\prod_{j=0}^\infty(1-x\q^j)\,.
\end{align}
Using this in the expression \eqref{legfprod} gives
\begin{align}
\frac{\gamma(z,m+r)}{\gamma(z,m)}=\frac{\theta(\EXP^{2\pi\ii(z+\sigma m)};\p\q)}{\theta(\EXP^{2\pi\ii(z-\tau m)};\p\q)}=(-1)^m\,\EXP^{-2\pi\ii m(z-\tau m)}\,(\p\q)^{-m(m-1)/2}\,,
\end{align}
where in the second step we used $z+\sigma m=(z-\tau m)+2\eta m$ and $\theta(x\q;\q)=-x^{-1}\theta(x;\q)$. A direct computation with \eqref{ellnorm} shows that $\phi_e(z,m+r)-\phi_e(z,m)$ is equal to minus the logarithm of the right-hand side, up to an integer multiple of $2\pi\ii$.  Note that $\gamma(z,m)$ is clearly $1$-periodic in $z$, and the phase in the last relation of \eqref{periodic} comes from the factor $\phi_e$. The second relation in \eqref{periodic} follows from the third, and if $r$ is odd, $m(r+1)$ is always even, and the lens elliptic gamma function is then $r$-periodic in $z$. 

For integer $m$, the poles of $\Gamma(z,m)$ lie in the closed lower half plane at the points
\begin{align}
\label{poles}
z=-(\sigma+\tau)\ell-\sigma(r\ell'+\bar m)+k\,,\qquad z=-(\sigma+\tau)\ell-\tau(r(\ell'+1)-\bar m)+k\,,
\end{align}
and the zeros lie in the open upper half plane at the points
\begin{align}
\label{zeros}
z=(\sigma+\tau)(\ell+1)+\sigma(r(\ell'+1)-\bar m)+k\,,\qquad z=(\sigma+\tau)(\ell+1)+\tau(r\ell'+\bar m)+k\,,
\end{align}
where $\ell,\ell'=0,1,\ldots$, $k\in\mathbb{Z}$, and $\bar m\in\{0,1,\ldots,r-1\}$ denotes $m$ mod $r$.   The set of zeros is $2\eta$ minus the set of poles of $\Gamma(z,-m)$, in agreement with the reflection formula \eqref{reflection}. %
The only poles on the real line are at integer points $z\in\mathbb{Z}$; they occur only for $\bar m=0$, and come from the first sequence in \eqref{poles} with $\ell=\ell'=0$. In particular, $\Gamma(z,0)$ has a simple pole at $z=0$. Since $\phi_e(z,0)=0$, we have
\begin{align}
\label{residue}
\lim_{z\to0}\frac{\Gamma(2z,0)}{\Gamma(z,0)}=\lim_{z\to 0}\frac{1-\EXP^{2\pi\ii z}}{1-\EXP^{4\pi\ii z}}=\frac12\,.
\end{align}

Finally, we write $\lambda$ for the constant
\begin{align}
\label{lambdadef}
\lambda=(\p^r;\p^r)_\infty\,(\q^r;\q^r)_\infty\,.%
\end{align}
We write collections of complex variables as vectors $\vect=(t_1,\ldots,t_{K})$, with $\eta-\vect\coloneq (\eta-t_1,\ldots,\eta-t_K)$, and similarly for integer variables, with $-\veca\coloneq (-a_1,\ldots,-a_K)$.

\paragraph{Contours.} All sum/integrals in this paper have the form
\begin{align}
\label{generalform}
\sum_{\vecy\in\mathbb{Z}_r^n}\int_{C^n}\prod_{i=1}^n\prod_{j}\Gamma(t_j\pm z_i,a_j\pm y_i)\times(\textrm{factors with no poles in } z_i)\prod_{i=1}^n dz_i\,,
\end{align}
with $\vecy=(y_1,\ldots,y_n)$ summing over $\{0,1,\ldots,r-1\}^n$. The integrand is $r$-periodic in each $y_i$ by \eqref{periodic}, and $1$-periodic in each $z_i$ provided $\sum_ja_j\equiv0$ (mod $r$): by the last relation in \eqref{periodic}, the shift $z_i\to z_i+1$ multiplies $\prod_j\Gamma(t_j\pm z_i,a_j\pm y_i)$ by $\EXP^{4\pi\ii y_i\sum_ja_j/r}$, and the other factors are invariant. By \eqref{poles}, the poles of $\Gamma(t_j+z_i,a_j+y_i)$ in $z_i$ lie at the points $-t_j$ plus points in the closed lower half plane, and those of $\Gamma(t_j-z_i,a_j-y_i)$ lie at $t_j$ plus points in the closed upper half plane. When $\im(t_j)>0$ for all $j$, the contour may be chosen to be $C=[0,1]$. Otherwise, for each $\vecy$, $C$ may be chosen to be a curve from $\ii c$ to $1+\ii c$ for some $c\in\mathbb{R}$ whose integer translates form a curve separating the poles of $\prod_j\Gamma(t_j+z,a_j+y_i)$ from those of $\prod_j\Gamma(t_j-z,a_j-y_i)$.  Note that the contour may depend on $y_i$. Such a contour exists whenever no pole of the first product coincides with a pole of the second, which is the case for generic $t_j$, and the sum/integral is then the analytic continuation in the $t_j$ of its value for $\im(t_j)>0$. In the limit that two such poles approach each other, the contour is pinched by those poles and the sum/integral in general becomes singular.

\section{\texorpdfstring{The $BC_n\leftrightarrow BC_{\tilde n}$ transformation and $\mathrm{Sp}(2n)$ duality}{The BCn <-> BCn' transformation and Sp(2n) duality}}\label{sec:bcn}

Let $n$ and $\tilde n$ be integers satisfying $n,\tilde n\geq0$, and let $\vect=(t_1,\ldots,t_{2\tilde n+2n+4})\in\mathbb{C}^{2\tilde n+2n+4}$ and $\veca=(a_1,\ldots,a_{2\tilde n+2n+4})\in\mathbb{Z}^{2\tilde n+2n+4}$ satisfy the balancing conditions
\begin{align}
\label{balancing2}
\sum_{j=1}^{2\tilde n+2n+4} t_j\equiv 2(\tilde n+1)\eta\quad (\textrm{mod }2r)\,,\qquad\sum_{j=1}^{2\tilde n+2n+4} a_j\equiv 0 \quad (\textrm{mod }r)\,.
\end{align}
The $BC_n$ sum/integral is defined by \cite{Kels:2017toi}
\begin{align}
\label{BCnIntDef}
I^{\tilde n}_{BC_n}(\vect,\veca)\coloneq \frac{\lambda^n}{2^n\,n!}\,\sum_{\vecy\in\mathbb{Z}_r^n}\,\int_{C^n}\frac{\prod_{i=1}^n\prod_{j=1}^{2\tilde n+2n+4}\Gamma(t_j\pm z_i,a_j\pm y_i)}{\prod_{i=1}^n\Gamma(\pm 2z_i,\pm 2y_i)\prod_{1\leq i<j\leq n}\Gamma(\pm z_i\pm z_j,\pm y_i\pm y_j)}\prod_{i=1}^ndz_i\,,
\end{align}
with $I^{\tilde n}_{BC_0}\coloneq 1$. The contour $C$ is as described at the end of Section~\ref{sec:defs}. The denominator does not produce poles: by \eqref{legfprod} and the reflection formula for $\Gamma_1$, $1/\Gamma(\pm z,\pm k)$ is, up to an exponential factor, a product of two theta functions, and hence an entire function of $z$ for every $k\in\mathbb{Z}$. The expression \eqref{BCnIntDef} is invariant under $t_j\to t_j+2r$ by \eqref{periodic}, so the balancing condition on the $t_j$ only needs to be considered modulo $2r$.

\begin{theorem}[\cite{Kels:2017toi}]\label{thm:bcn}
Under the balancing condition \eqref{balancing2},
\begin{align}
\label{transdef2}
I^{\tilde n}_{BC_n}(\vect,\veca)=\prod_{1\leq i<j\leq 2\tilde n+2n+4}\Gamma(t_i+t_j,a_i+a_j)\;I^n_{BC_{\tilde n}}(\eta-\vect,-\veca)\,.
\end{align}
\end{theorem}

For $r=1$ this is equivalent to Rains' $BC_n$ transformation formula \cite{Rains:2005lkc}. For $\tilde n=0$ it is a $BC_n$ sum/integral evaluation formula which was obtained in \cite{Spiridonov:2016uae}, and for $n=1$ it is the elliptic beta sum/integral formula \cite{Kels:2015bda}.

\paragraph{$\mathrm{Sp}(2n)$ duality.} Theorem~\ref{thm:bcn} is the equality of lens indices for the $\mathrm{Sp}(2n)$ Seiberg duality of Intriligator and Pouliot \cite{Intriligator:1995ne}. Consider $\mathrm{Sp}(2n)$ gauge theory with $2F=2(\tilde n+n+2)$ chiral multiplets $Q_j$ in the fundamental representation, with no superpotential. The anomaly-free R-charge is $R_Q=1-(n+1)/F$, and we parametrise
\begin{align}
t_j=R_Q\, \eta+\xi_j\,,\qquad \sum_{j=1}^{2F}\xi_j=0\,,
\end{align}
where $\xi_j$ are chemical potentials for the $\mathrm{SU}(2F)$ flavour symmetry and $a_j$ are the holonomies of the flavour background on $S^3/\mathbb{Z}_r$. Then \eqref{balancing2} holds, and $I^{\tilde n}_{BC_n}(\vect,\veca)$ is the lens index of this theory \cite{Benini:2011nc,Razamat:2013opa,Kels:2017toi}. The right-hand side of \eqref{transdef2} is the lens index of the magnetic theory. It has gauge group $\mathrm{Sp}(2\tilde n)=\mathrm{Sp}(2(F-n-2))$, $2F$ dual quarks with parameters $(\eta-t_j,-a_j)$, and mesons $M_{ij}=Q_iQ_j$ in the antisymmetric representation of $\mathrm{SU}(2F)$, which give the factor $\prod_{i<j}\Gamma(t_i+t_j,a_i+a_j)$.

\section{A duplication formula for the lens elliptic gamma function}\label{sec:dup}

Define the two integers $\epsilon_r$ and $\kappa_r$ by
\begin{align}
\label{kappadef}
\epsilon_r\coloneq r\text{ mod }2\,,\qquad \kappa_r\coloneq r^2-\epsilon_r\,,
\end{align}
and the constant
\begin{align}
\label{Ddef}
D_r\coloneq \EXP^{\pi\ii\kappa_r \eta/(2r)}\,.
\end{align}
We consider pairs $(u,b)\in\mathbb{C}\times\mathbb{Z}$, and define $\cS_r$ to be the following set of eight pairs
\begin{align}
\label{Sdef}
\cS_r\coloneq 
\begin{cases}
\cS\cup \left\{(0,\tfrac{r}{2}),\,(\tfrac12,\tfrac{r}{2}),\,(\eta,\tfrac{r}{2}),\,(\eta-\tfrac12,\tfrac{r}{2})\right\}, & r \textrm{ even}\,,\\
\cS\cup \left\{(\tfrac{\sigma}{2},\tfrac{r-1}{2}),\,(\tfrac{\sigma+1}{2},\tfrac{r-1}{2}),\,(\tfrac{\tau}{2},\tfrac{r+1}{2}),\,(\tfrac{\tau-1}{2},\tfrac{r+1}{2})\right\}, & r \textrm{ odd}\,,\\
\end{cases}
\end{align}
where
\begin{align}
\cS=\left\{(0,0),\,(\tfrac12,0),\,(\eta,0),\,(\eta-\tfrac12,0)\right\}.
\end{align}
When $r=1$, for each pair $(u,b)\in\cS_1$, we have $b\equiv 0$ mod $r$, and the eight distinct values of $\EXP^{2\pi\ii u}$ are $\pm1$, $\pm(\p\q)^{1/2}$, $\pm\p^{1/2}$, $\pm\q^{1/2}$, which correspond to the parameters used in \cite{Spiridonov:2011hf}.

The set $\cS_r$ has two simple properties:
\begin{align}
\label{Sprops}
\sum_{(u,b)\in\cS_r} u=4\eta\,,\qquad \sum_{(u,b)\in\cS_r} b=2r\equiv 0\quad(\textrm{mod }r)\,.
\end{align}
It is also invariant under the involution
\begin{align}
\label{involution}
\iota:\ (u,b)\mapsto (\eta-u,-b)\,,
\end{align}
when $b$ is taken modulo $r$. Indeed, $\iota$ exchanges $(0,0)\leftrightarrow(\eta,0)$ and $(\tfrac12,0)\leftrightarrow(\eta-\tfrac12,0)$, and similarly for the elements with $b=\tfrac{r}{2}$ ($r$ even). For $r$ odd, it exchanges $(\tfrac\sigma2,\tfrac{r-1}{2})\leftrightarrow(\tfrac\tau2,\tfrac{r+1}{2})$ and $(\tfrac{\sigma+1}2,\tfrac{r-1}{2})\leftrightarrow(\tfrac{\tau-1}2,\tfrac{r+1}{2})$. No element of $\cS_r$ is fixed by $\iota$. Since $\Gamma$ is $r$-periodic in its integer argument, $\Gamma(u+z,b+y)$ depends on $b$ only modulo $r$, and this allows us to freely use $\iota$ on products of such factors over $\cS_r$.

\begin{lemma}[Duplication formula]\label{lem:dup}
For $z\in\mathbb{C}$ and $y\in\mathbb{Z}$,
\begin{align}
\label{dup}
\prod_{(u,b)\in\cS_r}\Gamma(z+u,y+b)=D_r\EXP^{-\pi\ii\kappa_rz/r-2\pi\ii\epsilon_r y/r}\,\Gamma(2z,2y)\,.
\end{align}
Consequently,
\begin{align}
\label{duppm}
\prod_{(u,b)\in\cS_r}\Gamma(u\pm z,b\pm y)=D_r^2\,\Gamma(\pm 2z,\pm 2y)\,.
\end{align}
\end{lemma}

\begin{proof}
First consider the product of the factor $\gamma$. By \eqref{legfprod},
\begin{align}
\gamma(z+u,y+b)=\Gamma_1\big(X+u+\sigma b;r\sigma,2\eta\big)\,\Gamma_1\big(X'+u+\tau(\tfrac r2-b);r\tau,2\eta\big)\,,
\end{align}
where $X=z+\sigma y$ and $X'=z+\tau(\tfrac r2-y)$. Likewise, $\gamma(2z,2y)=\Gamma_1(2X;r\sigma,2\eta)\,\Gamma_1(2X';r\tau,2\eta)$. The elliptic gamma function satisfies the duplication formula
\begin{align}
\label{dupGamma1}
&\Gamma_1(2X;\omega_1,\omega_2)=\prod_{v\in E(\omega_1,\omega_2)}\Gamma_1(X+v;\omega_1,\omega_2)\,,%
\end{align}
where
\begin{align}
&E(\omega_1,\omega_2)\coloneq \left\{0,\tfrac12,\tfrac{\omega_1}2,\tfrac{\omega_1+1}{2},\tfrac{\omega_2}2,\tfrac{\omega_2+1}{2},\tfrac{\omega_1+\omega_2}2,\tfrac{\omega_1+\omega_2+1}{2}\right\},
\end{align}
which follows directly from the product formula \eqref{egf2}.  Indeed, writing $x=\EXP^{2\pi\ii X}$, $\p_1=\EXP^{2\pi\ii\omega_1}$, $\p_2=\EXP^{2\pi\ii\omega_2}$, the eight distinct values of $\EXP^{2\pi\ii v}$, where $v\in E(\omega_1,\omega_2)$, are $\pm1$, $\pm\p_1^{1/2}$, $\pm\p_2^{1/2}$, $\pm(\p_1\p_2)^{1/2}$, and thus $\prod_{v\in E}(1-x\EXP^{2\pi\ii v}\p_1^j\p_2^k)=\prod_{\alpha,\beta\in\{0,1\}}(1-x^2\p_1^{2j+\alpha}\p_2^{2k+\beta})$. 

Since $\Gamma_1$ is $1$-periodic, we consider elements of $E$ modulo $1$. Direct inspection of \eqref{Sdef} shows that, for both parities of $r$, the set $\{u+\sigma b\mid (u,b)\in\cS_r\}$ coincides with $E(r\sigma,2\eta)$ mod $1$. For example, for $r$ odd, $\tfrac\sigma2+\sigma\tfrac{r-1}{2}=\tfrac{r\sigma}{2}$ and $\tfrac\tau2+\sigma\tfrac{r+1}2=\tfrac{r\sigma}{2}+\eta$. Similarly, the set $\{u+\tau(\tfrac r2-b)\mid (u,b)\in\cS_r\}$ coincides with $E(r\tau,2\eta)$ mod $1$; e.g., for $r$ odd, $\tfrac\sigma2+\tau\big(\tfrac r2-\tfrac{r-1}{2}\big)=\eta$ and $\tfrac\tau2+\tau\big(\tfrac r2-\tfrac{r+1}{2}\big)=0$. Hence
\begin{align}
\prod_{(u,b)\in\cS_r}\gamma(z+u,y+b)=\gamma(2z,2y)\,.
\end{align}
It remains to compare the exponential factors. Since $\phi_e(z,m)$ is a polynomial in $z$ and $m$, an elementary computation shows that
\begin{align}
\label{phisum}
\sum_{(u,b)\in\cS_r}\phi_e(z+u,y+b)-\phi_e(2z,2y)=\frac{\pi\ii\kappa_r(\eta-2z)}{2r}-\frac{2\pi\ii\epsilon_r y}{r}\,,
\end{align}
as an identity of polynomials in $z$, $y$, $\sigma$ and $\tau$. This proves \eqref{dup}. Note that for $r$ odd the $y$-dependent term $-2\pi\ii y/r$ is not periodic in $y$ with period $r$ as a complex number, but $\EXP^{-2\pi\ii y/r}$ is, consistently with \eqref{periodic}. Multiplying \eqref{dup} by the same formula with $(z,y)\to(-z,-y)$ gives \eqref{duppm}: the $z$- and $y$-dependent terms in the exponent cancel, and the remaining constant is $2\left(\pi\ii\kappa_r \eta/(2r)\right)$.
\end{proof}

\begin{lemma}\label{lem:aux}
The following five identities hold:
\begin{align}
\label{aux1}
&\Gamma(\eta\pm z,\pm y)=1\,,\qquad 
\Gamma(\eta,0)=1\,,%
\end{align}
\begin{align}
\label{aux2}
&\prod_{\{(u,b),(u',b')\}\subset\cS_r}\Gamma(u+u',b+b')=1\,,%
\end{align}
\begin{align}
\label{aux3}
&\prod_{(u,b)\in\cS_r\setminus\{(0,0)\}}\Gamma(u,b)=\frac{D_r}{2}\,,\qquad
\prod_{(u,b)\in\cS_r\setminus\{(\eta,0)\}}\Gamma(\eta+u,b)=\frac{2}{D_r}\,.
\end{align}
\end{lemma}

\begin{proof}
The first identity in \eqref{aux1} is simply the reflection formula \eqref{reflection} with $z\to \eta+z$. The second identity in \eqref{aux1} follows from $\phi_e(\eta,0)=0$ and noting that in \eqref{littlegamma} at $(z,m)=(\eta,0)$ the products over $k$ telescope and it reduces to
\begin{align}
\gamma(\eta,0)=\prod_{j=0}^\infty\frac{1}{1-(\p\q)^{j+\frac12}}\,\left(1-(\p\q)^{j+\frac12}\right)=1\,.
\end{align}
For \eqref{aux2}, partition the $28$ unordered pairs $\{(u,b),(u',b')\}$ into orbits of $\iota$. There are four orbits of size one, of the form $\{(u,b),\iota(u,b)\}$, with each contributing a factor $\Gamma(\eta,0)=1$. The remaining $24$ pairs form orbits of size two, of the form $\{\{(u,b),(u',b')\},\{\iota(u,b),\iota(u',b')\}\}$, and the contribution from the products coming from each of these orbits is also $1$.  This follows from the reflection formula \eqref{reflection}, since the two pairs in such an orbit give the arguments $(u+u',b+b')$ and $(2\eta-u-u',-b-b')$.

For the first identity in \eqref{aux3}, set $y=0$ in \eqref{dup}, divide both sides by the factor $\Gamma(z,0)$ coming from $(u,b)=(0,0)$, and take $z\to0$ using \eqref{residue}. For the second, %
we can use the reciprocal of the first identity along with the reflection formula \eqref{reflection} and the involution $\iota$ as follows
\begin{align}
\frac{2}{D_r}=\!\!\prod_{(u,b)\in\cS_r\setminus\{(0,0)\}}\frac{1}{\Gamma(u,b)} 
=\!\!\prod_{(u,b)\in\cS_r\setminus\{(0,0)\}}\Gamma(2\eta-u,-b) %
=\!\!\prod_{(u,b)\in\cS_r\setminus\{(\eta,0)\}}\Gamma(\eta+u,b)\,,
\end{align}
where in the last step we used that $\iota:(u,b)\mapsto(\eta-u,-b)$ maps $\cS_r\setminus\{(0,0)\}$ to $\cS_r\setminus\{(\eta,0)\}$, with $b$ taken modulo $r$. This is the second identity in \eqref{aux3}.

\end{proof}

\section{\texorpdfstring{$\mathrm{SO}(N)$ sum/integrals}{SO(N) sum/integrals}}\label{sec:so}

\subsection{Definitions}

Let $N$ and $N_f$ be integers $N,N_f\geq 0$, and let $\vect=(t_1,\ldots,t_{N_f})\in\mathbb{C}^{N_f}$ and $\veca=(a_1,\ldots,a_{N_f})\in\mathbb{Z}^{N_f}$. We write $N=2n+\delta$ with $\delta\in\{0,1\}$.

\begin{definition}\label{def:so}
For $n\geq1$, the $\mathrm{SO}(2n)$ and $\mathrm{SO}(2n+1)$ sum/integrals are
\begin{align}
\label{SOeven}
I_{\mathrm{SO}(2n)}(\vect,\veca)&\coloneq \frac{\lambda^n}{2^{n-1}\,n!}\sum_{\vecy\in\mathbb{Z}_r^n}\int_{C^n}\frac{\prod_{i=1}^n\prod_{j=1}^{N_f}\Gamma(t_j\pm z_i,a_j\pm y_i)}{\prod_{1\leq i<j\leq n}\Gamma(\pm z_i\pm z_j,\pm y_i\pm y_j)}\prod_{i=1}^ndz_i\,,%
\end{align}
\begin{align}
\label{SOodd}
I_{\mathrm{SO}(2n+1)}(\vect,\veca)&\coloneq\frac{\lambda^n}{2^{n}\,n!}\prod_{j=1}^{N_f}\Gamma(t_j,a_j)\nonumber\\
&\quad\times\sum_{\vecy\in\mathbb{Z}_r^n}\int_{C^n}\frac{\prod_{i=1}^n\prod_{j=1}^{N_f}\Gamma(t_j\pm z_i,a_j\pm y_i)}{\prod_{i=1}^n\Gamma(\pm z_i,\pm y_i)\prod_{1\leq i<j\leq n}\Gamma(\pm z_i\pm z_j,\pm y_i\pm y_j)}\prod_{i=1}^ndz_i\,.
\end{align}
For $N=0,1$ we set
\begin{align}
\label{SOsmall}
I_{\mathrm{SO}(0)}(\vect,\veca)\coloneq2\,,\qquad I_{\mathrm{SO}(1)}(\vect,\veca)\coloneq\prod_{j=1}^{N_f}\Gamma(t_j,a_j)\,.
\end{align}
The contour $C$ is as described at the end of Section~\ref{sec:defs}; in particular we may choose $C=[0,1]$ when $\im(t_j)>0$ for all $j$.
\end{definition}

The prefactors $2^{n-1}n!$ and $2^nn!$ are the orders of the Weyl groups of $D_n$ and $B_n$. The integrand of \eqref{SOeven} is built from the weights $\pm e_i$ of the vector representation of $\mathrm{SO}(2n)$ and its roots $\pm e_i\pm e_j$. The integrand of \eqref{SOodd} also contains the zero weight of the vector representation of $\mathrm{SO}(2n+1)$, which gives $\prod_j\Gamma(t_j,a_j)$, and the short roots $\pm e_i$. The values \eqref{SOsmall} are chosen so that the main theorem below also holds at the boundary cases $N\in\{0,1\}$ or $\tN\in\{0,1\}$ (we will also see they are forced by \eqref{lhsred} below with $n=0$). The factor $2$ for $\mathrm{SO}(0)$ reflects the ratio $|W(B_n)|/|W(D_n)|=2$, which appears in the reduction from $BC_n$ to $D_n$. For $r=1$, \eqref{SOeven} and \eqref{SOodd} reduce to the $\mathrm{SO}(N)$ integrals of \cite[Section 2]{Spiridonov:2011hf}.

\subsection{Main theorem}

\begin{theorem}\label{thm:main}
Let $N$ and $\tN$ be integers satisfying $N,\tN\geq0$ and let $N_f\coloneq N+\tN-4$.  %
Assume that either $N_f\geq3$, or $N_f=2$ and $N$ is odd, and that $\vect\in\mathbb{C}^{N_f}$ and $\veca\in\mathbb{Z}^{N_f}$ satisfy
\begin{align}
\label{SObalancing}
\sum_{j=1}^{N_f}t_j\equiv (N_f-N+2)\,\eta\quad(\textrm{mod }2r)\,,\qquad \sum_{j=1}^{N_f}a_j\equiv 0\quad(\textrm{mod }r)\,.
\end{align}
Then, as an identity of meromorphic functions of $\vect$ subject to \eqref{SObalancing},
\begin{align}
\label{SOtrans}
I_{\mathrm{SO}(N)}(\vect,\veca)=\prod_{1\leq i\leq j\leq N_f}\Gamma(t_i+t_j,a_i+a_j)\;I_{\mathrm{SO}(\tN)}(\eta-\vect,-\veca)\,.
\end{align}
\end{theorem}

\begin{remark}
The product in \eqref{SOtrans} includes the diagonal terms $i=j$. The parameters $(\eta-\vect,-\veca)$ satisfy \eqref{SObalancing} with $N$ replaced by $\tN$. Since $\Gamma(t_i+t_j,a_i+a_j)\Gamma(2\eta-t_i-t_j,-a_i-a_j)=1$, the identity \eqref{SOtrans} is symmetric under the component-wise action of $\iota$ on $(\vect,\veca)$, with $N\leftrightarrow\tN$.    
\end{remark}

\begin{remark}
The restriction on $N_f$ is made because for certain small $N_f$ the balancing condition forces a degenerate situation. For $N_f=0$ it requires $N=\tN=2$, where \eqref{SOtrans} is trivial. For $N_f=1$, $N=0,\ldots,5$, it forces $2t_1\equiv(6-2N)\eta$ with $a_1\equiv0$, so that the factor $\Gamma(2t_1,2a_1)$ sits at a pole or a zero. The same happens for $\Gamma(t_1+t_2,a_1+a_2)$ when $N_f=2$ and $N$ is even ($N=0,2,4,6$). %
\end{remark}

\begin{proof}

Write $N=2n+\delta$ and $\tN=2\tilde n+\tilde\delta$ with $\delta,\tilde\delta\in\{0,1\}$, so that $N_f+8-\delta-\tilde\delta=2\tilde n+2n+4$. Let $\cS_{\delta,\tilde\delta}\subset\cS_r$ be the set obtained from $\cS_r$ by omitting $(0,0)$ if $\delta=1$ and omitting $(\eta,0)$ if $\tilde\delta=1$. By \eqref{involution},
\begin{align}
\iota(\cS_{\delta,\tilde\delta})=\cS_{\tilde\delta,\delta}\,.
\end{align}
Let $\vecu,\vecb$ be the vectors of the $8-\delta-\tilde\delta$ elements of $\cS_{\delta,\tilde\delta}$. We apply Theorem~\ref{thm:bcn} with parameters
\begin{align}
\vect_{BC}=(\vect,\vecu)\,,\qquad \veca_{BC}=(\veca,\vecb)\,.
\end{align}
By \eqref{Sprops}, we have
\begin{align}
\sum_{(u,b)\in\cS_{\delta,\tilde\delta}} u=(4-\tilde\delta)\eta\,,\qquad \sum_{(u,b)\in\cS_{\delta,\tilde\delta}}b\equiv0\; (\text{mod } r).
\end{align}
The balancing condition \eqref{balancing2} is therefore equivalent to \eqref{SObalancing}, since $2(\tilde n+1)-(4-\tilde\delta)=\tN-2=N_f-N+2$.

The special values $u\in\{0,\tfrac12\}$ are real. For the corresponding pairs $(u,b)\in\cS_{\delta,\tilde\delta}$, we have $b\in\{0,\tfrac r2\}$, and the poles of $\Gamma(u+z,b+y)$ and $\Gamma(u-z,b-y)$ meet at the real line at $z\equiv u$ (mod $1$) when $b+y\equiv0$ (mod $r$), and the contour is pinched on the left-hand side of \eqref{transdef2}.  Similarly, $\eta-u$ is real for the special values $u\in\{\eta,\eta-\tfrac12\}$ and this leads to poles of $\Gamma(\eta-u+z,-b+y)$ and $\Gamma(\eta-u-z,-b-y)$ pinching the contour on the right-hand side of \eqref{transdef2} at $z\equiv\eta-u$ (mod $1$) when $-b-y\equiv0$ (mod $r$). We therefore define the specialisation as a limit. 

Replace $u$ by $u+\ii\varepsilon$ for $u\in\{0,\tfrac12\}$ and by $u-\ii\varepsilon$ for $u\in\{\eta,\eta-\tfrac12\}$, with $\varepsilon>0$, and if $\delta\neq\tilde\delta$ shift $t_1$ by $\pm\ii\varepsilon$ to preserve the balancing condition. All other special values $u$ and $\eta-u$ have positive imaginary part. Hence, for generic $\vect$, Theorem~\ref{thm:bcn} applies for every small $\varepsilon>0$, with contours that do not depend on $\varepsilon$ near the real axis. On the left-hand side of \eqref{transdef2} we have pointwise convergence of
\begin{align}
\label{lhsspec}
\lim_{\varepsilon\to0^+}\frac{\prod_{(u,b)\in\cS_{\delta,\tilde\delta}}\Gamma(u\pm z,b\pm y)}{\Gamma(\pm2z,\pm2y)}=\frac{D_r^2}{\Gamma(\pm z,\pm y)^{\delta}}\,,
\end{align}
on the contour $C$ and away from the pinch points at $z_0\in\{0,\tfrac12\}$ for $2y_0\equiv0$ mod $r$.  The expression on the right-hand side of \eqref{lhsspec} follows from \eqref{duppm} and \eqref{aux1}.  The product on the left of \eqref{lhsspec} is also bounded uniformly in $\varepsilon$: near a pinch point each pinching pair contributes a factor of order $1/((z-z_0)^2+\varepsilon^2)$, and at most one such pair occurs for given $(z_0,y_0)$, while $1/\Gamma(\pm2z,\pm2y)$ has a double zero at $z_0$ independent of $\varepsilon$. By dominated convergence, the limit $\varepsilon\to0$ may be taken inside the integrals on both sides of \eqref{transdef2}.

Comparing normalisations on the left-hand side of \eqref{transdef2} we get %
\begin{align}
\label{lhsred}
I^{\tilde n}_{BC_n}(\vect_{BC},\veca_{BC})=2^{\delta-1}\,D_r^{2n}\,P(\vect,\veca)^{-\delta}\,I_{\mathrm{SO}(N)}(\vect,\veca)\,,\qquad P(\vect,\veca)\coloneq \prod_{j=1}^{N_f}\Gamma(t_j,a_j)\,.
\end{align}
This also holds for $n=0$ by \eqref{SOsmall}.  Note that since $\iota(\cS_{\delta,\tilde\delta})=\cS_{\tilde\delta,\delta}$, the special values on the right-hand side of \eqref{transdef2} are the elements of $\cS_r$ with $(0,0)$ removed if $\tilde\delta=1$ and $(\eta,0)$ removed if $\delta=1$. Thus, a similar limiting argument as above for the right-hand side gives
\begin{align}
\label{rhsred}
I^n_{BC_{\tilde n}}(\eta-\vect_{BC},-\veca_{BC})=2^{\tilde\delta-1}\,D_r^{2\tilde n}\,P(\eta-\vect,-\veca)^{-\tilde\delta}\,I_{\mathrm{SO}(\tN)}(\eta-\vect,-\veca)\,.
\end{align}

Finally, we split the product in \eqref{transdef2} into three parts.
\begin{itemize}
\item \emph{Pairs among the special values.} By \eqref{aux2}, \eqref{aux3}, and $\Gamma(\eta,0)=1$,
\begin{align}
\prod_{\{(u,b),(u',b')\}\subset\cS_{\delta,\tilde\delta}}\Gamma(u+u',b+b')=\left(\frac{D_r}{2}\right)^{-\delta}\left(\frac{2}{D_r}\right)^{-\tilde\delta}.
\end{align}
Here the product is over pairs of elements of $\cS_{\delta,\tilde\delta}$, which may be obtained from \eqref{aux2} by dividing out the pairs containing a removed element, using \eqref{aux3}. When $\delta=\tilde\delta=1$, the pair $\{(0,0),(\eta,0)\}$ is divided out twice, so we multiply back by $\Gamma(\eta,0)=1$.
\item \emph{Mixed pairs.} By \eqref{dup}, \eqref{aux1}, and \eqref{reflection},
\begin{align}
\begin{split}
\prod_{j=1}^{N_f}\prod_{(u,b)\in\cS_{\delta,\tilde\delta}}\Gamma(t_j+u,a_j+b)&=\prod_{j=1}^{N_f}\frac{D_r\EXP^{-\pi\ii\kappa_rt_j/r-2\pi\ii\epsilon_r a_j/r}\,\Gamma(2t_j,2a_j)}{\Gamma(t_j,a_j)^\delta\,\Gamma(\eta+t_j,a_j)^{\tilde\delta}}\\
&=D_r^{N-\tN}\,\frac{P(\eta-\vect,-\veca)^{\tilde\delta}}{P(\vect,\veca)^{\delta}}\prod_{j=1}^{N_f}\Gamma(2t_j,2a_j)\,.
\end{split}
\end{align}
In the last step we used $1/\Gamma(\eta+t_j,a_j)=\Gamma(\eta-t_j,-a_j)$ and the balancing condition \eqref{SObalancing}; $\sum_{j=1}^{N_f} a_j\equiv0$ (mod $r$) leads to the term involving $\epsilon_r$ vanishing, and $\sum_{j=1}^{N_f} t_j\equiv(\tN-2)\eta$ (mod $2r$) gives $\prod_{j=1}^{N_f}D_r\EXP^{-\pi\ii\kappa_rt_j/r}=D_r^{N_f-2\tN+4}=D_r^{N-\tN}$. %
\item \emph{Pairs among the $t_j$.} These give $\prod_{1\leq i<j\leq N_f}\Gamma(t_i+t_j,a_i+a_j)$.
\end{itemize}
Substituting all of this into \eqref{transdef2}, and using $D_r^{2n}=D_r^{N-\delta}$ and $D_r^{2\tilde n}=D_r^{\tN-\tilde\delta}$, every power of $D_r$, every power of $2$, and every factor of $P$ cancels. This leaves \eqref{SOtrans}.
\end{proof}

\begin{remark}
For $r=1$, Theorem~\ref{thm:main} reduces to the relations between the $\mathrm{SO}(N)$ indices and the $\mathrm{Sp}(2n)$ indices given in \cite[Section 2]{Spiridonov:2011hf}. In particular, the sets $\cS_{\delta,\tilde\delta}$ reduce to the parameter sets used there. For $r>1$, the choice \eqref{Sdef} matters. The $\gamma$-part of the proof of Lemma~\ref{lem:dup} determines the $u$ only modulo $1$, where $(u,b)\in\cS_{\delta,\tilde\delta}$. However, by the last relation in \eqref{periodic}, shifting a single $u$ by $1$, where $(u,b)\in\cS_{\delta,\tilde\delta}$ multiplies $\Gamma(u\pm z,b\pm y)$ by $\EXP^{2\pi\ii(b^2+y^2)/r}$, which depends on $y$ for $r>1$, and it also changes $\sum_{(u,b)\in\cS_{\delta,\tilde\delta}} u$ and hence the balancing condition. The choice \eqref{Sdef} satisfies \eqref{Sprops} and is $\iota$-invariant. As a result \eqref{duppm} contains no $y$-dependent phase, \eqref{aux2} holds, and the final formula \eqref{SOtrans} contains no extra phases.
\end{remark}

\section{Special cases and physical interpretation}\label{sec:phys}

\subsection{Evaluation formulas}

When $\tN\in\{0,1\}$, the right-hand side of \eqref{SOtrans} contains no integral, and Theorem~\ref{thm:main} reduces to an evaluation formula.

\begin{cor}\label{cor:eval}
\mbox{}
\begin{enumerate}
\item ($N_f=N-4$) Let $N$ be an integer $N\geq7$ and suppose that $\vect\in\mathbb{C}^{N-4}$ and $\veca\in\mathbb{Z}^{N-4}$ satisfy
\begin{align}
\sum_{j=1}^{N-4} t_j\equiv-2\eta\quad(\text{mod } 2r)\,,\qquad \sum_{j=1}^{N-4}a_j\equiv0\quad (\text{mod }r)\,.
\end{align}
Then
\begin{align}
\label{evalNm4}
I_{\mathrm{SO}(N)}(\vect,\veca)=2\prod_{1\leq i\leq j\leq N-4}\Gamma(t_i+t_j,a_i+a_j)\,.
\end{align}
\item  ($N_f=N-3$) Let $N$ be an integer $N\geq5$ and suppose that $\vect\in\mathbb{C}^{N-3}$ and $\veca\in\mathbb{Z}^{N-3}$ satisfy
\begin{align}
\sum_{j=1}^{N-3} t_j\equiv-\eta\quad(\text{mod } 2r)\,,\qquad \sum_{j=1}^{N-3}a_j\equiv0\quad (\text{mod }r)\,.
\end{align}
Then
\begin{align}
\label{evalNm3}
I_{\mathrm{SO}(N)}(\vect,\veca)=\prod_{1\leq i\leq j\leq N-3}\Gamma(t_i+t_j,a_i+a_j)\prod_{j=1}^{N-3}\Gamma(\eta-t_j,-a_j)\,.
\end{align}
\end{enumerate}
\end{cor}

For smaller $N$ the balancing condition either cannot be satisfied or leads to the degenerate cases discussed after Theorem~\ref{thm:main}. In Corollary \ref{cor:eval} the balancing condition requires $\im\big(\sum_{j}t_j\big)<0$, so the contour $C$ must be deformed off the real line as described in Section~\ref{sec:defs}. For $r=1$ these reduce to the $\mathrm{SO}(N)$ evaluation formulas of \cite[eqs.~(2.11)--(2.14)]{Spiridonov:2011hf}. The factor $2$ in \eqref{evalNm4} does not appear in \cite[eqs.~(2.13) and (2.14)]{Spiridonov:2011hf}. However, it follows from the reduction formulas \cite[eqs.~(2.7) and (2.8)]{Spiridonov:2011hf} and the $BC_n$ evaluation formula, exactly as in the proof of Theorem~\ref{thm:main}.

Physically, \eqref{evalNm3} corresponds to $\mathrm{SO}(N)$ SQCD with $N_f=N-3$ flavours. Its magnetic description has the trivial gauge group ``$\mathrm{SO}(1)$'', and consists of the mesons $M$ and the singlets $q$ with superpotential $\mathcal{W}=Mqq$, which describe one of the branches of the theory found in \cite{Intriligator:1995id}. The factor $2$ in \eqref{evalNm4} comes from the normalisation $I_{\mathrm{SO}(0)}=2$, that is, from the ratio $|W(B_n)|/|W(D_n)|=2$.

\subsection{Seiberg duality on the lens space}

Consider 4d $\mathcal{N}=1$ $\mathrm{SO}(N)$ gauge theory with $N_f$ chiral multiplets $Q_j$ in the vector representation and no superpotential \cite{Intriligator:1995id}. The anomaly-free R-charge is $R_Q=1-(N-2)/N_f$. As in Section~\ref{sec:bcn}, we parametrise
\begin{align}
t_j=R_Q\,\eta+\xi_j\,,\qquad \sum_{j=1}^{N_f}\xi_j=0\,,
\end{align}
with $\xi_j$ the $\mathrm{SU}(N_f)$ chemical potentials and $a_j$ the flavour holonomies, so that \eqref{SObalancing} holds. The sum over $\vecy\in\mathbb{Z}_r^n$ is the sum over flat $\mathrm{SO}(N)$ connections on $S^3/\mathbb{Z}_r$, whose holonomies may be conjugated into the maximal torus, and the integral over $\vecz$ implements the projection onto gauge invariant states. For odd $r$, the stabiliser in $\mathrm{SO}(N)$ of any such holonomy is a product of unitary groups and one special orthogonal group, which is connected, and $I_{\mathrm{SO}(N)}(\vect,\veca)$ is the full lens index of this theory on $S^1\times S^3/\mathbb{Z}_r$. For even $r$ it is the part of the lens index discussed below.

The magnetic theory \cite{Intriligator:1995id} has gauge group $\mathrm{SO}(\tN)$ with $\tN=N_f-N+4$, $N_f$ dual quarks $q^j$ with R-charge $1-R_Q$ in the conjugate flavour representation, and gauge singlet mesons $M_{ij}=Q_iQ_j$ in the symmetric tensor representation of $\mathrm{SU}(N_f)$, with superpotential $\mathcal{W}=Mqq$. Their contributions are exactly the factors $I_{\mathrm{SO}(\tN)}(\eta-\vect,-\veca)$ and $\prod_{i\leq j}\Gamma(t_i+t_j,a_i+a_j)$ in \eqref{SOtrans}. For odd $r$, Theorem~\ref{thm:main} is therefore the equality of the electric and magnetic lens indices, for all four combinations of parities of $N$ and $\tN$. The identity holds uniformly in $N$ and $\tN$, including the small values of $\tN$ for which the magnetic theory has special features \cite{Intriligator:1995id}.

For even $r$, the situation is richer. Following \cite{Razamat:2013opa}, the flat $\mathrm{SO}(N)$ bundles on $S^1\times S^3/\mathbb{Z}_r$ are labelled by their second Stiefel--Whitney class in $H^2(S^1\times S^3/\mathbb{Z}_r,\mathbb{Z}_2)\cong\mathbb{Z}_2\oplus\mathbb{Z}_2$, which we write as a pair of signs $(\mu,\nu)\in\{\pm1\}^2$. Here $\mu=1$ if the lift $\hat g\in \mathrm{Spin}(N)$ of the holonomy $g$ around the non-contractible cycle of $S^3/\mathbb{Z}_r$ satisfies $\hat g^r=1$, and $\mu=-1$ if $\hat g^r=-1$; in our variables, $\mu=(-1)^{\sum_iy_i}$. Similarly, $\nu=1$ if the lifts of the holonomies around the two cycles commute in $\mathrm{Spin}(N)$, and $\nu=-1$ if they anticommute. In the latter case the two holonomies cannot lie in a common maximal torus; these sectors come from the non-identity components of the stabiliser of $g$, which is disconnected. Denoting by $Z_{\mu,\nu}$ the contributions of the four sectors, normalised as in \cite[eqs.~(2.32) and (4.2)]{Razamat:2013opa}, the $\mathrm{Spin}(N)$ index is $Z_{1,1}$, the $\mathrm{SO}(N)_+$ index is $\frac12\sum_{\mu,\nu}Z_{\mu,\nu}$, and the $\mathrm{SO}(N)_-$ index is obtained from the $\mathrm{SO}(N)_+$ index by flipping the sign of $Z_{1,-1}$ for $r\equiv2$ (mod $4$) and of $Z_{-1,-1}$ for $r\equiv0$ (mod $4$) \cite{Aharony:2013hda,Razamat:2013opa}.

Our sum/integrals contain only the sectors with $\nu=1$, and
\begin{align}
\label{IZrel}
I_{\mathrm{SO}(N)}(\vect,\veca)=\tfrac12\big(Z_{1,1}+Z_{-1,1}\big)\,.
\end{align}
To see this, compare the weights of a given holonomy $g$ (in the maximal torus) on both sides. In the explicit expressions \cite[eqs.~(4.8) and (4.12)]{Razamat:2013opa}, the sum runs over Weyl-inequivalent $g$. The integral over the remaining torus variables uses the Haar measure of the identity component $Z_g^0$ of the centraliser of $g$, normalised by $1/|W(Z_g^0)|$. There is also a factor $\hat n=2$ if the two lifts $\pm\hat g\in\mathrm{Spin}(N)$ are not related by the Weyl group $W$ of $\mathrm{SO}(N)$, and $\hat n=1$ otherwise. In \eqref{SOeven} and \eqref{SOodd}, on the other hand, the sum over $\vecy\in\mathbb{Z}_r^n$ with the prefactor $1/|W|$ gives each Weyl orbit the weight $1/|W_g|$, where $W_g\subset W$ is the stabiliser of $g$. Since $\mathrm{Spin}(N)$ is simply connected and $\hat g$ has finite order ($\hat g^{2r}=1$), hence is semisimple, the centraliser of $\hat g$ in $\mathrm{Spin}(N)$ is connected by Steinberg's theorem \cite[Theorem~8.1]{MR230728}. Its Weyl group is therefore the subgroup of $W_g$ that fixes $\hat g$, and this subgroup coincides with $W(Z_g^0)$. The other elements of $W_g$ map $\hat g$ to $-\hat g$. Hence $|W_g|=(2/\hat n)\,|W(Z_g^0)|$, so $1/|W_g|$ equals $\tfrac12\hat n/|W(Z_g^0)|$, which is the weight of $g$ in $\frac12(Z_{1,1}+Z_{-1,1})$. The integrands coincide, which proves \eqref{IZrel}. For example, for $\mathrm{SO}(3)$ and $r=2$, the holonomy $y=0$ has $\hat n=2$ and $W_g=W$, while $y=1$ has $\hat n=1$ and $W(Z_g^0)$ trivial.

Theorem~\ref{thm:main} therefore shows that the $\nu=1$ parts of the electric and magnetic lens indices agree. This is consistent with the relations between the sectors of dual theories observed in \cite[eqs.~(4.5) and (4.7)]{Razamat:2013opa} by expanding the lens index. Both sets of relations imply that $Z_{1,1}+Z_{-1,1}$ of the electric theory equals the corresponding combination of the magnetic theory, including the mesons. Theorem~\ref{thm:main} proves this equality exactly, for all even $r$. For odd $r$, $H^2(S^1\times S^3/\mathbb{Z}_r,\mathbb{Z}_2)=0$, there is only one sector, and all global forms give the same index. Correspondingly, there is no difference between even and odd $r$ in the final formula \eqref{SOtrans}, although the intermediate duplication formula depends on the parity of $r$.

\section{Concluding remarks}\label{sec:concl}

We have derived sum/integral transformation formulas for the orthogonal groups. Together with \cite{Kels:2017toi}, they give sum/integral identities for the lens index Seiberg dualities of all classical gauge groups: $\mathrm{SU}(N)$, $\mathrm{Sp}(2N)$ and $\mathrm{SO}(N)$. We close with some open problems.

For even $r$, the most immediate open problem is to include the sectors with $\nu=-1$. In these sectors the holonomies can be brought to a block form with a fixed $3\times3$ block and a remaining $\mathrm{SO}(N-3)$ part \cite{Razamat:2013opa}, so their contributions are again sum/integrals of lower rank with some parameters fixed to special values. Together with Theorem~\ref{thm:main}, sum/integral identities for these sectors would give the equality of the full lens indices of $\mathrm{SO}(N)_\pm$ and $\mathrm{Spin}(N)$ theories, and test the expected dualities $\mathrm{Spin}(N)\leftrightarrow \mathrm{SO}(\tN)_-$ and $\mathrm{SO}(N)_+\leftrightarrow \mathrm{SO}(\tN)_+$ \cite{Aharony:2013hda,Razamat:2013opa}. We have not been able to obtain them from Theorem~\ref{thm:bcn} by the method of this paper.

For $r=1$, many other orthogonal group dualities, for example those with tensor matter, can be obtained by similar reductions of $\mathrm{Sp}$ identities \cite{Spiridonov:2011hf}. Lemma~\ref{lem:dup} should make the lens space versions of these reductions equally direct. Dualities with spinor matter require new ingredients.

The hyperbolic, trigonometric and rational limits of elliptic hypergeometric integrals give identities for 3d theories \cite{Dolan:2011rp,Gahramanov:2016wxi}, and the analogous limits of the lens sum/integrals give identities for 3d theories on lens spaces \cite{Imamura:2012rq} and their degenerations \cite{Gahramanov:2016ilb,Sarkissian:2018ppc}. The limits of Theorem~\ref{thm:main} should similarly give orthogonal group identities, related to the 3d dualities for $\mathrm{SO}(N)$, $\mathrm{O}(N)_\pm$ and $\mathrm{Spin}(N)$ gauge groups \cite{Aharony:2013kma}; see also \cite[Section~4.3]{Lee:2016zud}, where the latter are obtained from the $\mathrm{SO}(N)$ duality by gauging discrete $\mathbb{Z}_2$ symmetries.

The $A_n$ transformation of \cite{Kels:2017toi} implies a star-star relation for an integrable lattice model \cite{Bazhanov:2013bh,Yamazaki:2013nra}, in line with the Gauge/YBE correspondence \cite{Yamazaki:2012cp,Terashima:2012cx,Yamazaki:2013nra} (see \cite{Yamazaki:2018xbx,Gahramanov:2017ysd,Gahramanov:2022qge} for reviews). The lens elliptic gamma function also gives solutions of the Yang--Baxter equation at roots of unity \cite{Kels:2017vbc,Yamazaki:2026kta}, and it has been used to construct lens generalisations of $\tau$-functions for the elliptic discrete Painlev\'e equation \cite{Kels:2018xnf}. Surface defects in the lens index act by elliptic difference operators, which are related to elliptic Ruijsenaars--Schneider operators and to transfer matrices of integrable spin chains \cite{Razamat:2013jxa,Maruyoshi:2026nrl,Yamazaki:2026kws}. It would be interesting to see whether the orthogonal group identities have a similar interpretation.

\section*{Acknowledgements}

This work was initiated when APK was at the University of Tokyo as an overseas researcher under the Postdoctoral Fellowship of the Japan Society for the Promotion of Science (JSPS).
The research of MY was supported in part by the World Premier International Research Center Initiative (WPI), MEXT, Japan; JSPS KAKENHI Grant No.~23K25865; JST, Japan (CREST Grant No.~JPMJCR26XA, Moonshot R\&D Grant No.~JPMJMS256E); and IBM-UTokyo-sponsored research.
The authors were assisted by Anthropic Claude and OpenAI ChatGPT in writing and editing the manuscript and in numerical checks.

\bibliographystyle{utphys}
\bibliography{main}

\end{document}